# EXPERIMENTAL EVIDENCE FOR OPTIMAL RESONANCES IN PION-NUCLEUS SCATTERING

D.B. Ion[1], M.L.D. Ion[2], R. Ion-Mihai[2], T.Angelescu[2] and A.A.Radu[3]

[1]*Institute for Physics and Nuclear Engineering, Department of Fundamental Physics, POB MG 6, Bucharest, Romania*
[2]*Bucharest University, Department for Nuclear Physics, POB MG 11, Bucharest, Romania*
[3] *Institute for Space Sciences, MG-23 Bucharest, Romania*

**Abstract:** Experimental data on the pion-nucleus total cross sections are analyzed in terms of the optimal resonances predictions (OR). The OR-predictions are found in a good agreement with the actual experimental data in the region corresponding to the $\Delta(1236)$ resonance in elementary pion-nucleon scattering. The essential signatures of the optimal resonances are experimentally verified with high accuracy.

Resonance is a fundamental paradigm in physical sciences and engineering. $\Delta(1236)$ resonance is a member of baryon decuplet, and exists in four electric charge states (isospin I=3/2) and has a total spin of $J = 3/2$. The $\Delta(1236)$ was first observed as a resonant interaction of a beam of π-mesons with a proton target. The probability of a scattering interaction between a pion and the proton is strongly dependent on energy, attaining a maximum at the $\Delta$-mass of $M_\Delta = 1236$ MeV. The $\Delta(1236)$ resonance plays an important role in a wide variety of nuclear phenomena, even under conditions of low energy and momentum transfer. Pion-nucleus interactions have been extensively investigated in the past three decades with the meson factories at LAMPF, TRIUMF, and PSI. The very extensive list of publications (see for example the review papers [1,2]) on pion-nucleus interactions cannot all be cited in this short article. We include only the references most closely related to our discussions here. The essential results obtained from the experimental data on pion-nucleus scattering, in the region corresponding to the $\Delta(1236)$ resonance in the elementary pion-nucleon interaction, are characterized by the particle-*wave duality* feature which includes: <u>*a resonant energy behavior*</u> manifested in the total, elastic and inelastic, cross sections (see refs. [3-11]), and <u>*a typical diffraction pattern*</u> observed in the pion-nucleus angular distributions [12-15](see also [37,38,41-76] in Ref. [1]). The theoretical papers on the pion-nucleus scattering in this energy region have achieved only a limited agreement between theory and experimental data. In Refs.[2] we introduced new exotic *collective states* of the nucleus called *dual diffractive resonances* (DDR) which can be excited in the hadron-nucleus interactions. The DDR description of pion-nucleus scattering, at the energies corresponding to the elementary $\pi N$ resonances, gives an interesting theoretical picture capable of quantitative results by which we can understand the above dual characteristic features.

In this paper the pion-nucleus scattering is described by the excitation of bound nucleons into the $\Delta(1236)$ resonance leading to optimal isobaric resonances of the whole nucleus. These resonances, called optimal resonances [see [16-17] are described according to the *principle of minimum distance in space of quantum states* (PMD-SQS) introduced in Ref. [16]. From the qualitatively good agreement with the data, we conclude that those PMD-SQS-giant isobaric resonances are a general feature of the nuclear excitation spectrum in the $\Delta(3,3)$ energy range. Therefore, for the nuclear part of the pion-nucleus scattering, we use the following expression:

$$f_N(E,x) = \frac{1}{2k}\sum(2l+1)\frac{\Gamma_{ell}}{E_{0l}-E-\frac{i}{2}[\Gamma_l-\gamma_0(E_{0l}-E)]}P_l(x) \quad (1)$$

where E is the c.m. energy, k is the c.m. momentum, $x \equiv \cos\theta$, $\theta$ - c.m. scattering angle. $E_{ol}, \Gamma_l, \Gamma_{ell}$, and $\gamma_{0l}$ are the effective parameters (mass, total width, elastic width and asymmetry parameter) of the "induced resonances" in each hadron-nucleus partial wave. The asymmetry parameter $\gamma_0$ is introduced in refs. [2] in a natural way starting with a Regge expression, $f_l(E) \propto [l-\alpha_l(E)]^{-1}$, for the pion-nucleus partial amplitude $f_l(E)$. So, in the neighborhood of the resonance energy $E_{0l}$ in which $\text{Re}\,\alpha(E_{0l})$ has an integer value, we have

$$f_l(E) = \frac{\Gamma_{ell}}{2k}[(E_{0l}-E)-\frac{1}{2}i(\Gamma_l-\gamma_{ol}(E_{0l}-E))]^{-1} \quad (2)$$

where $\Gamma_l \equiv 2\,\text{Im}\,\alpha(E_{0l})/\text{Re}\,\dot\alpha(E_{0l})$ and $\gamma_{0l} \equiv 2\,\text{Im}\,\dot\alpha(E_{0l})/\text{Re}\,\dot\alpha(E_{0l})$. So, the asymmetry parameter $\gamma_{0l}$ is a dimensionless parameter.

Now, using the principle of minimum distance in space of quantum states (see Refs. [16-17]) with the unidirectional constraint $\frac{d\sigma}{d\Omega}(E,1) = fixed$ we obtain

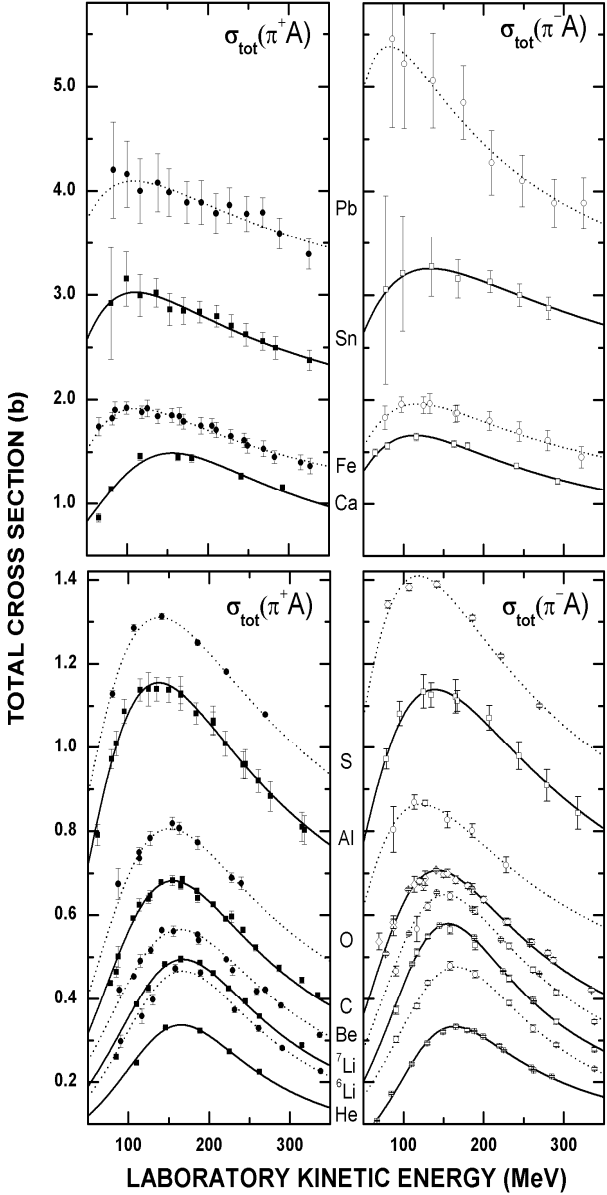

**Fig. 1**: The pion-nucleus total cross sections in the $\Delta(1236)$ - energy region. The experimental data are from Refs. [3-11]. The curves are results of the best fit to the data with the optimal resonance predictions Eq. (6).

$$f_N(E,x) = f(E,1)\frac{K_{L_o}(x,1)}{K_{L_o}(1,1)} \quad (3)$$

Where $K_{L_o}(x,y), x,y \in [-1,+1]$, are reproducing kernels of the Hilbert space of states [16]. In our particular case for the optimal resonances in the pion-nucleus scattering on spinless nuclei we have

$$2K(x,1) = \sum_{l=0}^{L_o}(2l+1)P_l(x)P_l(1) = \dot{P}_{L_o+1}(x) + \dot{P}_{L_o}(x) \quad (4)$$

$$2K(1,1) = \Sigma_{l=0}^{L_o}(2l+1) = (L_o+1)^2 = \frac{4\pi}{\sigma_{el}}\frac{d\sigma}{d\Omega}(E,1) \quad (5)$$

where $x = \cos\theta$, $P_l(x)$ and $P_l(y)$ are Legendre polynomials and $\dot{P}_l(x) \equiv \frac{dP_l(x)}{dx}$.

In this case, in the optimal resonance limit [17], the hadron-nucleus scattering is characterized by the following essential features:

(i) The energy behavior of the total hadron-nucleus cross section is of the asymmetric Breit-Wigner form given by

$$\sigma_T = \pi\lambdabar^2(L_o+1)^2 \frac{\Gamma_{el}[\Gamma - \gamma_0(E_0-E)]}{(E_0-E)^2 + \frac{1}{4}[\Gamma - \gamma_0(E_0-E)]^2} \quad (6)$$

$$\sigma_{el} = \pi\lambdabar^2(L_o+1)^2 \frac{\Gamma_{el}^2}{(E_0-E)^2 + \frac{1}{4}[\Gamma - \gamma_0(E_0-E)]^2} \quad (7)$$

(ii) The real part of the forward hadron-nucleus scattering amplitude has a resonant behavior described by

$$\mathrm{Re}f_{\pi A}(E,0^\circ) = \frac{\lambdabar(L_o+1)^2}{2} \frac{\Gamma_{el}(E_0-E)}{(E_0-E)^2 + \frac{1}{4}[\Gamma - \gamma_0(E_0-E)]^2} \quad (8)$$

(iii) The angular distributions of the optimal resonances are typical diffractive patterns, very sensitive to the values of the optimal angular momentum $L_o$. They are described by

$$\frac{d\sigma}{d\Omega}(E,x) = \frac{d\sigma}{d\Omega}(E,1)\left[\frac{K_{L_o}(x,1)}{K_{L_o}(1,1)}\right]^2 \quad (9)$$

The number of the maxima and minima in the entire $[-1,+1]$-interval are given by $N_{max} = L_o+1$ and $N_{min} = L_o$.

(iv) The optimal resonances saturate the "axiomatic" bounds:

$$\sigma_T^2(E) \leq 4\pi\lambdabar^2(L_o+1)^2\sigma_{el}(E) \quad (10)$$

$$\Gamma_n \leq \Gamma_1 n, \; n \equiv L_o+1 \quad (11)$$

at the energy $E = E_0$, where $\lambdabar = 1/k$.

Now we examine the energy behavior of the experimental data [3-15] on the pion-nucleus scattering in the region corresponding to the $\Delta(1236)$ resonance in the elementary pion-nucleon scattering. In applications of the optimal resonance mechanism for pion-nucleus scattering at energies where the elementary $\Delta(1236)$ resonances are produced, $\Gamma_1$ from Eq. (11) is identified with the $\Delta - total\ width$. Moreover, if the optimal angular momentum is taken as $L_o \approx kR$ where R ($R \approx r_0 A^{1/3}$) is the equivalent spherical radius of a nucleus with A-nucleons, then in all Eqs (1)-(11) we can write

$$\lambdabar^2 (L_o + 1)^2 \cong (R + \lambdabar)^2 \qquad (12)$$

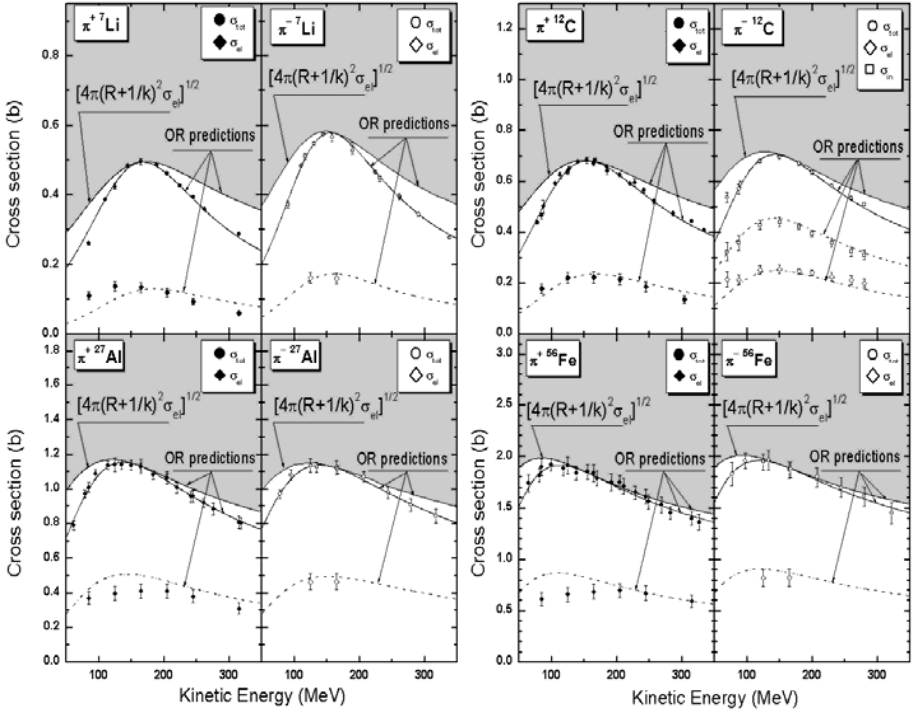

**Fig. 2**: The pion-nucleus total cross sections in the $\Delta(1236)$-energy region. The experimental data are from Ref. [3-11]. The curves are results of the best fit to the data with the optimal resonance predictions, Eq. (6). The saturation of the axiomatic optimal bound (13) is experimentally evident (see the text).

Consequently, at optimal hadron-nucleus resonance, from (10)-(11) we have the saturation of the following important bounds

$$\sigma_T^2(E) \leq 4\pi (R + \lambdabar)^2 \sigma_{el}(E) \qquad (13)$$
$$\Gamma_\Delta \leq \Gamma_{\pi A} \leq \Gamma_\Delta A^{1/3} \qquad (14)$$

at the energy $E = E_0$.

Now, the experimental data on the pion-nucleus total cross sections [3-11] are presented in Figs. 1 and 2. We see that the cross sections for all nuclei exhibit a maximum clearly related to $\Delta(1236)$ resonance, which shifts downward and broadens with increasing A and has an asymmetry which also increases with increasing A. It is easy to see that all these characteristic features are fairly reproduced by the optimal resonance predictions.

In fitting Eq. (6) to the experimental data we have considered the $E_0$ fixed by the relation

$$E_o = M_A + 1236 MeV - m_N, \quad \Gamma_{el} = k\gamma_1, \quad L_o \approx kR$$

and the geometric radius R fixed as in the tables 1 and 2 of Ref. [2], for each nucleus. The other parameters $\gamma_0, \gamma_1$ and $\Gamma$ from Eq. (4) are allowed to vary for each nucleus in order to obtain the best $\chi^2$-fit of the total cross sections. The optimal resonance parameters are presented in Fig.3.

Therefore, from the results presented in Figs. 1 we see that all experimental data on the pion-nucleus total cross sections are well described by Eq. (6). Moreover, by using the fitted parameters of the total cross sections we obtain numerical predictions (see Fig.2) for $\sigma_{el}, \sigma_{in}$ as well as for the upper bound $[4\pi(R + \lambdabar)^2 \sigma_{el}]^{1/2}$ which is saturated at the optimal resonance energy $E = E_0$. Then, we see that all optimal resonance predictions (see Figs. 1-2) are in excellent agreement with the available experimental data [3-11].

In conclusion in this paper we analyzed the pion-nucleus scattering data from the point of view of the excitation of the collective optimal resonances of the nucleus. The results and conclusions may be summarized as follows:

(i) The pion-nucleus total cross sections, in the energy region corresponding to $\Delta(1236)$ resonance in the elementary pion-nucleon interaction, are well described by the predictions of the optimal resonances dominance (see Figs. 1 and 2).

(ii) The total widths, obtained by fit of the total cross sections in the $\Delta(1236)$ energy region, are consistent with $\Gamma_{\pi A} = \Gamma_\Delta A^{1/3}$ for $\Gamma_\Delta = (120 \pm 5)$ MeV.

(iii) The available experimental data on $\sigma_{el}, \sigma_{in}$ are also in good agreement with the predictions of the PMD-SQS-resonance mechanism (see figs. 1, 2) but more experimental data are required.

(iv) The saturations of the axiomatic bound (13)-(14), at the optimal resonance energy $E = E_0$, are verified experimentally with high accuracy (see Figs. 2-3).

(v) Extensive data on pion-nucleus elastic scattering from a wide range of nuclei, obtained at meson factories (see Refs.[37, 38, 41–78] in Ref.[1]), all show the characteristics of diffractive scattering. That is, minima in the elastic scattering angular distributions correspond to the diffractive minima produced by strong optical absorption. This experimental result is also one of the important signatures of the optimal resonance dominance given in Eq. (9). We note of course that, in order to confirm the dominance of the optimal resonances hypothesis, in pion-nucleus scattering in the elementary resonance region, a quantitative analysis of the experimental data on the angular distributions is also necessary.

other than pion-nucleus scattering (e.g. electron-nucleus, nucleon-nucleus, nucleus-nucleus, etc.) can excite the same optimal resonance (or DDR states [2]). For example, in the nuclear photo excitation by electron scattering $^{12}C(e, e')$ a peak in the $\Delta(1236)$ energy region is clearly observed [18]. Detailed investigations on the nature of such phenomena are of great theoretical interest since they might provide strong evidence for the *universality* of the *optimal resonances spectrum*.

A detailed paper on the optimal resonances analysis in pion-nucleus scattering is now in preparation.

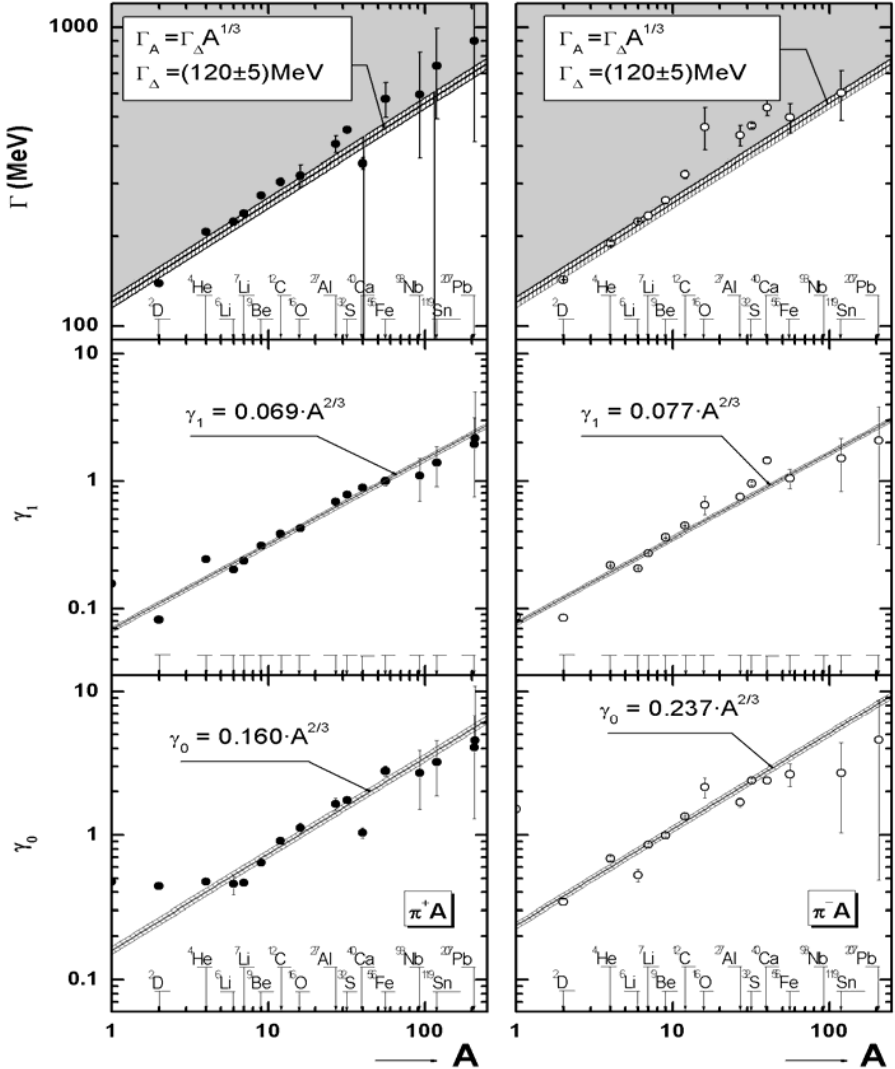

**Fig. 3**: The parameters $\gamma_0$, $\gamma_1$, and $\Gamma$ of Eq. (6) obtained by minimum $\chi^2$-fits to the total pion-nucleus cross sections. The solid circles are obtained from the fits to $\pi^+ - nucleus$ data, while the open circles are from fits to $\pi^- - nucleus$ data. The solid lines represent the smooth A-dependence of the corresponding parameter. The saturation of the axiomatic optimal bound (14) is experimentally verified with high accuracy (see the text).

On the other hand, one of the immediate consequences of the *optimal resonance concept* [17] is that reactions

### References

[1] T.-S.H.Lee and R.P.Redwine, Annu. Rev. Nucl. Part. Sci. **52** (2002), 23–63 and there in references
[2] D.B.Ion and Reveica Ion-Mihai, Nucl. Phys. **A 360,** 400 (1981) and the quoted references
[3] M.L. Scott et al.*,* Phys. Rev. Lett. **28**, 1209 (1972)
[4] A.S. Clough et al., Nucl. Phys. **B 76,** 15 (1974)
[5] C. Wilkin et al.*,* Nucl. Phys. **B 62,** 61 (1973)
[6] D. Ashery et al., Phys. Rev. **C 23**, 2173 (1981)
[7] F. Binon et al., Nucl. Phys. **B17**, 168 (1970); Nucl. Phys. **B33,** 42 (1971); Nucl. Phys. **B40**, 608 (1972)
[8] A.S. Carroll et al., Phys. Rev. **C 14,** 635 (1976)
[9] B.W. Allardice et al., Nucl. Phys. **A209,** 1 (1973)
[10] K. Junker et al., Phys. Rev. **C 43**, 1911 (1991)
[11] E. Pedroni, et al., Nucl. Phys. **A300,** 321 (1978)
[12] F. Binon et al., Nucl. Phys. **A298** (1978) 499
[13] J. Piffaretti et al. Phys. Lett. **71B,** 324 (1977)
[14] L.E. Antonuk et al. Nucl. Phys. **A420**,43 (1984)
[15] M.J. Leitch et al. Phys. Rev. *C* **29**, 561 (1984)
[16] D.B.Ion, Phys. Lett. **B 376** (1996) 282.
[17] D.B. Ion and M.L.D. Ion, Rom. Rep. Phys. **59**, 1058 (2007)
[18] F.H. Heimlich *et al.* Nucl. Phys. A231 (1974) 509